\documentclass[12pt]{article}
\newcommand{\R}{{\bf R}}
\newcommand{\C}{{\bf C}}
\newcommand{\quat}{{\bf H}}
\newcommand{\grad}{{\rm grad}}
\newcommand{\curl}{{\rm curl}}
\newcommand{\Z}{{\bf Z}}
\newcommand{\x}{\underline{x}}

\begin{document}

\title{\vspace{-1in}\parbox{\linewidth}{\small\hfill\shortstack{UCLA/99/TEP/23}}
\vspace{0.6in}\\ Supergravity Solution for\\ M5-brane Intersection}
\author{Sergey A. Cherkis\thanks{Research supported in part by Department of Energy under Grant \#FG03-91\-ER40662, Task C}\\
{\sl\small TEP, Department of Physics}\\
{\sl\small University of California, Los Angeles}\\
{\sl\small CA 90095-1547 USA}\\
{\sl\small cherkis@physics.ucla.edu}}

\begin{titlepage}
\renewcommand{\thepage}{ }
\renewcommand{\today}{ }
\thispagestyle{empty}

\maketitle

\begin{abstract}
Supergravity solution describing two intersecting M5-branes is presented. 
The branes are fixed in the relative transverse directions and are delocalized along the overall transverse ones. The intersection can be smoothed, so that 
the M5-branes present one holomorphic cycle. We also obtain a solution
corresponding to an M5-brane on a holomorphic cycle of multi-Taub-NUT space.
All these solutions preserve 1/4 of supersymmetry.
\end{abstract}

\end{titlepage}

\section{Introduction}
Besides the problem of finding supersymmetric solutions of supergravity being interesting
in itself, such solutions can be very useful. Since branes can be thought of as solitons of supergravity,
corresponding solutions can provide valuable information about their 
world--volume theories. Especially interesting would be solutions with branes wrapped on holomorphic cycles, branes ending on branes and intersecting branes.

After few years of attempts to obtain a supergravity solution corresponding to
a pair of intersecting branes \cite{PT, AT, GKT, GGPT, HYang, DYoum, MP, FS, GKMT}, finding such a solution of eleven-dimentional supergravity
explicitly with all branes fully localized is still an open problem. In \cite{FS}
this problem was reduced to solving a nonlinear equation\footnote{Solution to this equation as a perturbative expansion is studied in
\cite{GKMT}.} for a K\"ahler metric
of the form 
\begin{equation}
\Delta_\perp g_{m\hat{n}}+2\partial_m\partial_{\hat{n}}\det g_{k\hat{l}}=0.
\end{equation}

The brane configurations that are known explicitly are of two kinds:
(i) those with at least one brane delocalized in its relative transverse directions
\cite{T2}
and (ii) one in the near horizon limit of a six-brane with another brane 
ending on it \cite{Aki} or localized inside \cite{ITY}. We should also mention 
a solution presented in \cite{GKT} with two M5-branes intersecting over a string,
with both branes localized in the relative transverse directions. Here we are going to present a somewhat 
complimentary solution to those mentioned above. In its simplest version it describes a pair of  inersecting
M theory 5-branes. These are localized in the relative transverse
directions, but are delocalized in the overall transverse directions $(4,5,6)$,
as represented by the following table
\begin{center}
\begin{tabular}{l|ccccccccccc} 
   & 0 &1 &2 &3 &4 &5 &6 &7 &8 &9 &10 \\ \hline
M5 & x &x &x &x &s &s &s &x &x &  &       \\
M5 & x &x &x &x &s &s &s &  &  &x  &x \\ 
\end{tabular}
\end{center}
where `x' denotes a direction along the worldvolume of a five-brane and
`s' a normal direction along which the five-brane is smeared. In terms of the complex 
coordinates $v=y_7+i y_8$ and $w=y_9+i y_{10}$ this M5-brane intersection 
is described by
\begin{equation}\label{singular}
v w=0.
\end{equation}
This configuration does not satisfy
the rules for localized intersections derived in \cite{ETT}, where a systematic 
study of localized intersections is performed. The ansatz of 
\cite{ETT} assumes some natural symmetries, such as rotational symmetry in the
(7,8) plane. We wold like 
to point out that our solution does not posess these symmetries. The reasons 
for this are geometrical and are similar to the ones that cause the solution of \cite{Aki} not to be rotationaly symmetric. 

In the framework we are using, one can also describe a resolution of the
singular intersection Eq.(\ref{singular}) to a one smooth five-brane localized at
\begin{equation}\label{vw}
v w=\epsilon.
\end{equation}
That is, the five-brane worldvolume is $\R^4\times\Sigma$ with $\R^4$ parallel
to the $(0,1,2,3)$ directions and $\Sigma$ given by Eq.~(\ref{vw}).
And the most general solution we can obtain describes M theory `compactified'
on Taub-NUT space with the $\Sigma$ part of the five-brane worldvolume wrapping
an infinite two-cycle of Taub-NUT or multi-Taub-NUT.

The answer is given in terms of a solution to a linear equation. Namely the
following Laplace equation in three dimensions
\begin{equation}
\partial_z\partial_{\bar{z}} H(z,x)+\Omega(z)\partial_x\partial_x H(z,x)=0, 
\end{equation}
where $\Omega$ is a particular function of $z$ and we allow delta function sources on the right-hand side.

Our construction is based on ideas presented in \cite{ITY} and \cite{Aki}. Section~2 starts with some mathematical trivia in order to set up our notation.
Then we briefly describe the construction of Hashimoto \cite{Aki}, and
how we reverse it to obtain our results. Section~3 consists of the chain of dualities
leading to the final answer. The last section contains our conclusions.

\section{Setting}
In this section we set up our notation, describe the construction of \cite{ITY, Aki}
and how we reverse it to obtain the intersecting M5-brane solution. 
\subsection{Notation}
Consider
a flat four-dimensional space $\R^4=\C^2$ with coordinates $(v,w)$
\begin{equation}
v=\rho e^{i\phi}\cos\theta, \ \ w=\rho e^{i\psi} \sin\theta.
\end{equation}
Thinking of $\C^2$ as a quaternionic space $\quat$ we form $q=v i+w j$. We can 
write $q$ as $q=a e^{i\phi}$, with $a$ being purely imaginary quaternion $(\bar{a}=-a)$,
and introduce coordinates $x_7, x_8, x_9, x_{10}$ so that
\begin{equation}\label{coordinates}
i x_7+j x_8+k x_9=a i \bar{a}, \ \ x_{10}=2 \phi.
\end{equation}
Explicitly
\begin{eqnarray}\nonumber \label{trans}
x_7&=&\rho^2 \cos(2\theta)\\
x_8&=&\rho^2 \cos(\psi+\phi)\sin(2\theta)\\
\nonumber
x_9&=&\rho^2 \sin(\psi+\phi)\sin(2\theta).
\end{eqnarray}
In terms of these new coordinates the flat metric $ds^2=4\left(dv d\bar{v}+dw d\bar{w}\right)$
is 
\begin{equation}\label{metric}
ds^2=\frac{1}{r} {d\vec{r}} ^2+r\left(d x_{10}+\vec{\omega}\cdot\vec{r}\right)^2,
\end{equation}
where $\vec{r}=(x_7,x_8,x_9)$ and $\grad\ (1/r)=\curl\ (\vec{\omega})$.

Geometrically the three-sphere for a fixed distance $\rho^2$  is a Hopf 
fibration with $(2\theta, \phi+\psi)$ being spherical coordinates of the base $S^2$ 
and $\phi$ a coordinate along the $S^1$ fiber. To be more precise these are the
coordinates of the trivialization above the upper patch of $S^2$ ($2\theta<\pi$).
In the lower patch ($2\theta>0$) the proper fiber coordinate is 
$x_{10}=2\phi-2(\psi+\phi)=-2\psi$.

We shall be interested in various holomorphic cycles in this $\C^2$ space, 
so let us consider how they look in the different coordinate systems.

A plane
\begin{equation}\label{Z}
\{v=0\}=\{\theta=\pi/2\}=\{x_8=x_9=0,\ x_7\leq 0\},
\end{equation}
an orthogonal plane
\begin{equation}
\{w=0\}=\{\theta=0\}=\{x_8=x_9=0,\ x_7\geq 0\},
\end{equation}
two intersecting planes
\begin{equation}\label{intersecting}
\{v w=0\}=\{\theta=\pi/2\}\cup\{\theta=0\}=\{x_8=x_9=0\},
\end{equation}
resolved intersection parameterized by $\epsilon=\mu e^{i \alpha}$
\begin{equation}
\{v w=\epsilon\}=\{\psi+\phi=\alpha, \rho^2\sin(2\theta)=2\mu\}=\{x_8=2\mu\cos\alpha,x_9=2\mu\sin\alpha\}.
\end{equation}
In terms of $(x_7,x_8,x_9)$ coordinates these are respectively 
\begin{description}
\item a semi-infinite line starting
from the origin, 
\item a semi-infinite line in the opposite direction, 
\item a line passing
through the origin 
\item a line missing the origin by a distance $2\mu$.
\end{description}

Taub-NUT metric is
\begin{equation} \label{TN}
ds^2=V(r) {d\vec{r}} ^2+V^{-1}(r)\left(d x_{10}+\vec{\omega}\cdot\vec{r}\right)^2,
\end{equation} 
with $x_{10}\sim x_{10}+4\pi$, $V(r)=1+1/r$ and $\grad\ V=\curl\ (\vec{\omega})$. Near $r=0$ it approaches the flat space metric of Eq.~(\ref{metric}). Degenerate multi-Taub-NUT is
a quotient of the above metric by $\Z_N: x_{10}\mapsto x_{10}+4\pi/N$.

\subsection{Hashimoto's Origami}
Let the coordinates of the eleven-dimensional space of M theory be $y_i,\ i=0,\ldots,6$ and $v=y_7+i y_8, w=y_9+i y_{10}$.
The solution obtained by Akikazu Hashimoto in \cite{Aki} starts with an M theory five-brane located 
at $y_4=y_5=y_6=0, v=0$. The corresponding supergravity solution is well known
and is invariant with respect to rotation $v\rightarrow e^{i\chi} v$ as well as
$w\rightarrow e^{i\chi} w$. In terms of the new coordinates (\ref{coordinates}), interpreting $x_{10}$ as a direction
along the M theory circle ,  this is a solution of 
type IIA supergravity. In the absence of the five-brane the resulting
metric (\ref{metric}) is that of a Taub-NUT space near the origin of the Taub-NUT. Thus in type IIA it is interpreted as a near horizon of a D6-brane located at the origin. 
As for the M5-brane, it wraps
the M theory circle, and thus appears (see (\ref{Z})) as a D4-brane ending on 
the D6-brane (in the near horizon limit of the latter).

One can also consider a quotient with respect to the $\Z_N$ action
$(v, w)\rightarrow (\exp(2\pi i/N) v, \exp(-2\pi i/N) w)$, as presented in \cite{Aki}. In this case
the metric (\ref{metric}) is that of a multi-Taub-NUT space with $N$
coincident centers (in the neighborhood of the centers). Thus the resulting 
IIA solution is that of a D4-brane ending on $N$ coincident D6-branes in the near-horizon 
limit of the D6-branes.

\subsection{Unfolding the Origami}
In Hashimoto's picture one M5-brane corresponds to a D4-brane ending on 
a D6-brane. Let us note, that a pair of intersecting M5-branes (located at
$z w=0$) according to (\ref{intersecting}) corresponds to a D4-brane piercing 
the D6-brane. Thus if we were to know the latter solution of IIA supergravity  (with both branes being localized), we 
would be able to reconstruct the eleven-dimensional solution with two 
intersecting M5-branes. To be more precise knowing the solution near the D6-brane allows one to construct two intersecting M5-branes in flat space.
The full solution gives an M5-brane on a cycle of Taub-NUT space
(given e.g. by $x_8=x_9=0$) pinched at the origin.

Knowing a solution with a D4-brane orthogonal to a D6-brane, with the D4
localized away from the D6 at some non-zero $(x_8, x_9)$, would allow one to write a solution
with an M5-brane on a smooth cycle of Taub-NUT. If we limit ourselves 
to the near-horizon geometry of the D6-brane, in M theory we obtain 
a smooth M5-brane on a cycle $v w=\epsilon$ in flat space. 

Let us also note that a set of
$N$ parallel D6-branes pierced by a D4-brane yields an M5-brane on a cycle of 
multi-Taub-NUT with $N$ centers.

Even though no explicit solution with both intersecting branes being localized is known,
we can consider one with a D6-brane localized and an intersecting D4-brane delocalized in the three directions parallel to the D6. Such a
solution is reinterpreted in M theory as a pair of intersecting M5-branes
delocalized in the three overall transverse directions $(y_4, y_5, y_6)$.
This situation is somewhat different from that considered in \cite{T2} and \cite{DYoum}, where the  answer is given in terms of two functions: one harmonic and
the other is harmonic in the background of the metric given by the first one.
Since the space transverse to the smeared D4-brane is only two-dimensional and
does not have a positive harmonic function on it
this recipe has to be modified.

\section{D4 Smeared Along D6}
In order to obtain this solution and check its supersymmetry we shall start with  a IIB solution with a D3-brane parallel to a D7-brane and T-dualize it. The T-duality is performed 
along the three directions parallel to the D7 and orthogonal to the D3-brane.
Thus the D3 becomes a D6-brane (fully localized), and a D7 becomes a D4 
delocalized along the three directions mentioned.
\begin{center}
\begin{tabular}{l|cccccccccc} 
   & 0 &1 &2 &3 &4 &5 &6 &7 &8 &9  \\ \hline
D7 & x &x &x &x &x &x &x &x &  &         \\
D3 & x &x &x &x &  &  &  &  &  &  \\ 
\end{tabular}

$|$

$T_{4 5 6}$

$\downarrow$

\begin{tabular}{l|cccccccccc} 
   & 0 &1 &2 &3 &4 &5 &6 &7 &8 &9  \\ \hline
D4 & x &x &x &x &s &s &s &x &  &         \\
D6 & x &x &x &x &x &x &x &  &  &  \\ 
\end{tabular}
\end{center}

\subsection{D3 parallel to D7}
Let us split the coordinates into those parallel to both branes (with indices 
$a,b=0,1,2,3$), relative transverse ($i,j=4,5,6,7$) and transverse to both
branes ($m,n=8,9$).
We start with the following ansatz (in the string frame), which is a modification of the solution of \cite{GSVY},
\begin{eqnarray}
\label{ansatz}\nonumber
&&ds^2=e^{\phi/2} H^{-1/2} \eta_{ab}dx^adx^b+
e^{\phi/2} H^{1/2}\left(\delta_{ij}dx^i dx^j+\Omega dz d\bar{z}\right),\\
\nonumber &&F_{0123j}=\partial_j H^{-1},\ F_{0123m}=\partial_m H^{-1},\\
&&F_{ijk89}=\epsilon_{ijkl} \Omega \partial_l H,\ F_{4567n}=-\epsilon_{nm}\partial_m H\\
\nonumber
&&\chi+i e^{-\phi}=\tau(z)
\end{eqnarray}
where $z=x_8+i x_9$, $H=H(x_i,z,\bar{z})$ and
\begin{equation}\label{hol}
j(\tau(z))=a+\frac{1}{z},\ \Omega=\tau_2(z)\eta^2(\tau)\overline{\eta^2(\tau)}(z\bar{z})^{-1/12}.
\end{equation}
The five-form $F$ is self-dual with respect to the metric in Eq.~(\ref{ansatz}). For  $F$ to be closed the function $H(x_i,z,\bar{z})$ has to satisfy
\begin{equation}\label{consistency}
\partial_z\partial_{\bar{z}} H+\Omega(z)\partial_i\partial_i H=0.
\end{equation}
The source of the four-form potential $A$ with the self-dual field strength 
$F$ is a D3-brane. Thus presence of a D3-brane corresponds to introducing 
delta function sources on the right-hand side of Eq.(\ref{consistency}).

This ansatz is a particular case of the one considered in \cite{Kehagias}, where
the supersymmetry conditions are also analyzed. Now we briefly state the 
results of \cite{Kehagias} specified to our case.

The supersymmetry transformations of dilatino $\lambda$ and gravitino 
$\psi_{\mu}$ can be found in \cite{Schwarz} (where the Einstein frame $ds_E^2=\exp(-\phi/2)ds^2$ is used).
In our case we have $\delta\lambda$ proportional to
\begin{equation}\label{lambda}
\gamma^\mu\partial_\mu\tau \epsilon,
\end{equation}
and $\delta\psi_\mu$ proportional to
\begin{equation}\label{delta}
D_\mu\epsilon+\frac{i}{480}\gamma^{\mu_1\ldots\mu_5}F_{\mu_1\ldots\mu_5}\gamma_\mu \epsilon,
\end{equation}
with $\epsilon$ being complex Weyl ($\gamma^{11}\epsilon=\epsilon$) and
\begin{equation}
D_\mu=\partial_\mu+\frac{1}{4}\omega_\mu^{AB}\gamma_{AB}+
\frac{i}{4}\frac{\partial_\mu\tau_1}{\tau_2}.
\end{equation}
Splitting the gamma matrices with respect to $SO(1,3)\times SO(6)\in SO(1,9)$
as $\gamma^{\mu}=\Gamma^{\mu}\otimes 1$ for $\mu=0,\ldots,3$ and $\gamma^{\mu}=\Gamma^{5}\otimes \Gamma^{\mu}$ for $\mu=4,\ldots,9$,
 decompose the spinor parameter $\epsilon=H^{-1/8}\theta\otimes\eta$ so that
\begin{equation}\label{projection}
i\Gamma^0\Gamma^1\Gamma^2\Gamma^3\theta=\theta,\ -i\Gamma^4\Gamma^5\Gamma^6\Gamma^7\Gamma^8\Gamma^9\eta=\eta.
\end{equation}
Now the conditions of supersymmetry invariance following from Eqs.(\ref{lambda}) and (\ref{delta}) reduce to
\begin{equation}\label{pz}
\Gamma^{\bar{z}}\eta=\left(\Gamma^8+i\Gamma^9\right)\eta=0
\end{equation}
and
\begin{equation}\label{susy}
\left(\partial_{\mu_1}+\frac{1}{4}\omega_{{\mu_1}\mu_2\mu_3}\Gamma^{\mu_2\mu_3}+\frac{i}{4}\frac{\partial_{\mu_1}\tau_1}{\tau_2}\right)\eta=0,\ \mu_i=4,\ldots,9,
\end{equation}
respectively.

The spin connection $\omega_{{\mu_1}\mu_2\mu_3}$ above is computed in the metric \begin{equation}
ds^2=\delta_{ij}dx^i dx^j+\Omega dz d\bar{z}
\end{equation}
inside the brackets in Eq.(\ref{ansatz}).
The integrability condition for Eq.(\ref{susy})
\begin{equation}
\partial_z\partial_{\bar{z}} H+\Omega(z)\partial_i\partial_i H=0
\end{equation}
is automatically satisfied by our ansatz Eqs.(\ref{ansatz}, \ref{hol}, \ref{consistency}).

Condition of Eq.(\ref{projection}) cuts supersymmetry in half. Even though there are two equations, the second is satisfied because of the Weyl condition. Condition
(\ref{pz}) breaks another half, so one quarter of the supersymmetries is preserved by our ansatz.

\subsection{T-dual Solution}
Applying T-duality to the IIB 
solution in directions (4,5,6) with $x_4\sim x_4+2\pi R_4, x_5\sim x_5+2\pi R_5, x_6\sim x_6+2\pi R_6$ (as in \cite{BHO})  we obtain
\begin{eqnarray}\nonumber
ds^2&=&e^{\phi/2} H^{-1/2} \eta_{ab}dx^a dx^b+e^{-1/2} H^{-1/2}\left(\frac{d\x_4^2}{R_4^2}+\frac{d\x_5^2}{R_5^2}
+\frac{d\x_6^2}{R_6^2}\right)\\
&&+e^{\phi/2}H^{1/2}H^{1/2}\left(dx_7+\Omega(z) dz d\bar{z}\right),\\
\phi_A&=&\frac{1}{4}\phi-\frac{3}{4}\log H-\log R_4R_5R_6,\\
A^{(1)}_\alpha&=&4 A_{\alpha 456},\\
C_{5\alpha\beta}&=&-\frac{8}{3}A_{46\alpha\beta},\ C_{6\alpha\beta}=\frac{8}{3}A_{4\alpha\beta 5},\
C_{456}=\frac{2}{3}\chi,
\end{eqnarray}
where $\x_l\sim\x_l+2\pi$, $l=4,5,6$ are periodic coordinates and $A$ is a 
four-form potential of the field strength $F$ of Eq.(\ref{ansatz}).

\subsection{M theory solution}
Lifting the above IIA solution to that of the eleven-dimensional supergravity  \cite{BHO} and denoting by $V$ the volume $R_4R_5R_6$ we have
\begin{eqnarray}\nonumber\label{finale}
ds_{11}^2&=V^{2/3}&\left(e^{\phi/3}\eta_{ab}dx^a dx^b+
e^{-2\phi/3}\left(\frac{d\x_4^2}{R_4^2}+\frac{d\x_5^2}{R_5^2}
+\frac{d\x_6^2}{R_6^2}\right)+\right.\\
\nonumber &&\left.+e^{\phi/3}H\left(dx_7^2+\Omega dz d\bar{z}\right)\right)\\
&+V^{-4/3}& e^{\phi/3}H^{-1}\left(dx_{10}+4 A_{\alpha 456} dx^{\alpha}\right)^2,\\ \nonumber
C_{456}&=\frac{2}{3}\chi,&\ C_{5\alpha\beta}=-\frac{8}{3}A_{46\alpha\beta},\ \ 
C_{6\alpha\beta}=\frac{8}{3}A_{4\alpha\beta 5},
\end{eqnarray}
$\alpha, \beta=7,8,9$. We recall that $\chi+i e^{-\phi}=\tau(z)$ and
\begin{eqnarray}\nonumber
d A&=&d H^{-1}\wedge dx^0\wedge dx^1\wedge dx^2\wedge dx^3\\
&&-\Omega \epsilon_{ijkl} \partial_i H dx^j\wedge dx^k\wedge dx^l\wedge dx^8\wedge dx^9\\
\nonumber &&-\epsilon_{nm} \partial_m H dx^4\wedge dx^5\wedge dx^6\wedge dx^7\wedge dx^n.
\end{eqnarray}

Transformation to conventional spherical coordinates in the $(7,8,9,10)$ space
is given by Eq.~(\ref{trans}). This solution implicitly depends on two choices. 
One is the meromorphic function of $z$ determining $j(\tau)$ in Eq.(\ref{hol}) and the 
other is positions of the sources of the function $H(x,z,\bar{z})$. Let us note that if we pick $j(\tau)$ to be constant, then $\Omega$ is also constant
and 
\begin{equation}
H(x,z,\bar{z})=H(\infty)+\sum_j\frac{1}{\sqrt{|z-z_j|^2+(x-x_j)^2/\Omega}}.
\end{equation}
Thus we recover a usual multi-Taub-NUT space. In general in Eq.~(\ref{finale})
$H$ defines the background geometry the five-branes live in. For example, in the case of $H(\infty)=0$ a change of the asymptotic conditions that $H$ satisfies corresponds to deforming the four-dimensional flat subspace of M theory space to Taub-NUT ($H(\infty)\neq0$). Allowing multiple sources for $H$ would produce
a multi-Taub-NUT space. The choice of $j(\tau)$ determines where the smeared 
five-branes are located as well as the regularization of the metric on the conical
space far away from them\footnote{A D7-brane cannot exist alone, as its dilaton field becomes infinite some finite distance from the brane. This problem is resolved by 
introducing nonperturbative seven-branes parallel to it with corresponding monodromies $S$ or $TS$ of $SL(2,\Z)$ for example. These are located at points
$z_*$ where $j(\tau(z_*))=0,1$. For similar reasons an M5-brane smeared along two directions has to be regularized, but we can choose the points $z_*$ to be far away from the five-branes.}. Thus moving the source in the Laplace equation away from the pole of $j(\tau)$ 
corresponds to deforming the intersection to a smooth curve.

Approximate solutions to Eq.(\ref{consistency}) can be found in \cite{AFM} and \cite{KPR}, where an elliptic fibration over the $z$ plane determined by $\tau$ is considered and Seiberg-Witten coordinates $a$ and $a_D$ are introduced so that
\begin{equation}
d a=\eta^2(\tau)z^{-1/12}d z,\ \ d a_D=\tau d a.
\end{equation} In terms of these coordinates
Eq.(\ref{consistency}) is
\begin{equation}
\partial_a\partial_{\bar{a}}H+\frac{1}{2i}\left(\frac{\partial a_D}{\partial a} -\frac{\partial{\bar{a}_D}}{\partial{\bar{a}}}\right)\partial^2_x H=0,
\end{equation}
and $H$ can be expanded in the distance away from a Taub-NUT center.

Also, near the five-brane (small $z$) $1/z=j(\tau)\approx e^{-2\pi i\tau}$ and 
$\eta(\tau)\approx e^{\pi i\tau/12}$. Thus 
Eq.(\ref{consistency}) becomes
\begin{equation}
\partial_z\partial_{\bar{z}}H-\frac{1}{2\pi}\log|z| \partial_x^2 H=0.
\end{equation}
In case the source of $H$ is at $z=0$, which corresponds to the unresolved intersection, it is natural to presume axial symmetry. Introducing a coordinate
$t$ such that $|z|=e^{-t}$, consider
a function $f(t,\omega)$ which tends to infinity at $t\rightarrow+\infty$ 
and satisfies
\begin{equation}
\partial_t^2 f(t,\omega)-\frac{\omega^2}{2\pi} t e^{-2t} f(t,\omega)=0.
\end{equation}
Then $H(x,z,\bar{z})$ is given by
\begin{equation}
H(x,z,\bar{z})=\int_0^\infty \cos(\omega x) f(-\log|z|,\omega)\frac{d\omega}{2\pi}.
\end{equation}

\section{Conclusions}
We have obtained a supergravity solution describing the intersection of two M theory five-branes. 
The branes are still not fully localized in the directions transverse to both
of them. However, since in our case the problem becomes linear, this solution might be of use in linearizing the equation of Fayyazuddin and Smith \cite{FS}.

Another possibility to fix the five-branes completely might be to start with 
a IIA solution with a D4 brane fixed and D6 delocalized along it. In this case
interpreting it in M theory we obtain a pair of intersecting five-branes on a singular space. One might hope
to find a limit of parameters where the singularity disappears, which would
provide the fully localized five-brane solution.

A different interpretation of this solution is as follows. After compactifying 
one of the directions the five-branes are smeared over as the M theory 
circle and T-dualizing the other two directions, one obtains an intersection of two D7-branes. Both seven-branes are completely fixed then. This intersection
can be smoothed or put on Taub-NUT space as well.

The supergravity solution given also describes a five-brane on a holomorphic cycle fixed in the directions with nontrivial geometry. We did not succeed in fixing the positions of the branes completely, but we hope 
that the solution we presented will help in reaching that goal.

\section*{Acknowledgments}
It is a pleasure to thank Elena C\'aceres and Eric D'Hoker for stimulating
conversations.

\end{document}